\documentclass[conference]{IEEEtran}
\ifCLASSINFOpdf
\else
\fi
\usepackage{graphicx}

\begin{document}
%
% paper title
% can use linebreaks \\ within to get better formatting as desired
\title{[CSE 5304 Fall 2012] \\ Study of Betweenness Centrality Algorithms \\in GPU}

% author names and affiliations
% use a multiple column layout for up to three different
% affiliations
\author{\IEEEauthorblockN{Saad Quader}
\IEEEauthorblockA{Department of Computer Science and Engineering\\
University of Connecticut\\
Storrs, CT-06226\\
Email: saad.quader@uconn.edu}
%\and
%\IEEEauthorblockN{Homer Simpson}
%\IEEEauthorblockA{Twentieth Century Fox\\
%Springfield, USA\\
%Email: homer@thesimpsons.com}
%\and
%\IEEEauthorblockN{James Kirk\\ and Montgomery Scott}
%\IEEEauthorblockA{Starfleet Academy\\
%San Francisco, California 96678-2391\\
%Telephone: (800) 555--1212\\
%Fax: (888) 555--1212}
}

% conference papers do not typically use \thanks and this command
% is locked out in conference mode. If really needed, such as for
% the acknowledgment of grants, issue a \IEEEoverridecommandlockouts
% after \documentclass

% for over three affiliations, or if they all won't fit within the width
% of the page, use this alternative format:
% 
%\author{\IEEEauthorblockN{Michael Shell\IEEEauthorrefmark{1},
%Homer Simpson\IEEEauthorrefmark{2},
%James Kirk\IEEEauthorrefmark{3}, 
%Montgomery Scott\IEEEauthorrefmark{3} and
%Eldon Tyrell\IEEEauthorrefmark{4}}
%\IEEEauthorblockA{\IEEEauthorrefmark{1}School of Electrical and Computer Engineering\\
%Georgia Institute of Technology,
%Atlanta, Georgia 30332--0250\\ Email: see http://www.michaelshell.org/contact.html}
%\IEEEauthorblockA{\IEEEauthorrefmark{2}Twentieth Century Fox, Springfield, USA\\
%Email: homer@thesimpsons.com}
%\IEEEauthorblockA{\IEEEauthorrefmark{3}Starfleet Academy, San Francisco, California 96678-2391\\
%Telephone: (800) 555--1212, Fax: (888) 555--1212}
%\IEEEauthorblockA{\IEEEauthorrefmark{4}Tyrell Inc., 123 Replicant Street, Los Angeles, California 90210--4321}}

% use for special paper notices
%\IEEEspecialpapernotice{(Invited Paper)}

% make the title area
\maketitle

\begin{abstract}
%\boldmath
The problem of computing the \textit{Betweenness Centrality} (BC) is important in analyzing graphs in many practical applications like social networks, biological networks, transportation networks, electrical circuits, etc. Since this problem is computation intensive, researchers have been developing algorithms using high performance computing resources like supercomputers, clusters, and Graphics Processing Units (GPUs). Current GPU algorithms for computing BC employ Brandes' sequential algorithm with different trade-offs for thread scheduling, data structures, and atomic operations. In this paper, we study three GPU algorithms for computing BC of unweighted, directed, scale-free networks. We discuss and measure the trade-offs of their design choices about balanced thread scheduling, atomic operations, synchronizations and latency hiding. Our program is written in NVIDIA CUDA C and was tested on an NVIDIA Tesla M2050 GPU.
\end{abstract}
% IEEEtran.cls defaults to using nonbold math in the Abstract.
% This preserves the distinction between vectors and scalars. However,
% if the conference you are submitting to favors bold math in the abstract,
% then you can use LaTeX's standard command \boldmath at the very start
% of the abstract to achieve this. Many IEEE journals/conferences frown on
% math in the abstract anyway.

% no keywords

% For peer review papers, you can put extra information on the cover
% page as needed:
% \ifCLASSOPTIONpeerreview
% \begin{center} \bfseries EDICS Category: 3-BBND \end{center}
% \fi
%
% For peerreview papers, this IEEEtran command inserts a page break and
% creates the second title. It will be ignored for other modes.
\IEEEpeerreviewmaketitle

\section{Introduction and Background}\label{sec:intro}

Betweenness centrality (BC) of a vertex in any network is the fraction of all pairwise shortest paths in the network that pass through that vertex. A vertex with high BC in a network holds an influential position in communication within the whole network. Therefore, BC of a vertex is an important metric in various network analysis application, for example: social network analysis, transportation network analysis, clustering, etc.

Computing BC involves the computation of all pair shortest paths (APSP) of the network. Suppose the network has $n$ vertices and $m$ edges. Then, the time and space complexity of the APSP problem, if done using Floyd-Warshall algorithm, is $O(n^3)$. Thus it becomes impractical for large networks where $n$ is large. For directed graph, Dijkstra's Single Source Shortest Path (SSSP) algorithm, if run from every vertex, can solve APSP in $O(mn)$ time and $O(n^2)$ space. For unweighted graphs, BFS can also do the same. The breakthrough came in 2001 when Brandes \cite{brandes} gave a faster sequential algorithm which exploits a recurrence relation in accumulating the shortest-path dependencies of all vertices. This algorithm still runs in $O(mn)$ time but in $O(n+m)$ space. A natural extension was to develop the parallel version of this algorithm on various platforms: Cray MTA-2 and Cray XMT \cite{bader_lock_free}, IBM Cyclops \cite{tan_bc_ibm}, etc. In \cite{bader_lock_free}, the authors showed that it is possible to eliminate the use of critical region while accumulating dependencies for vertices. In another vein, since computing exact BC is computation heavy, algorithms for computing approximate BC on high-performance platforms were also proposed \cite{bader_approx}. However, these were supercomputers and hence not available to most researchers.

With the advent of GPU, massively parallel computing was brought to the reach of many researchers and naturally many graph algorithms, such as APSP, was implemented in GPU \cite{harish_apsp_cuda}. The first BC algorithm on GPU was implemented by Sriram et al. \cite{sriram_bc}. They used adjacency lists to represent the graph, and used one thread block for each source node and explored all nodes at the BFS frontier (that is, vertices at the same level) in parallel. Therefore this algorithm exploits parallelism in two levels: high level and mid level. But the drawback of this approach is that all neighbors of a vertex are explored sequentially, which means threads responsible for high-degree nodes takes more time than other threads. This causes load-imbalance in the GPU and thus hurts the performance if the graph has a power-law degree distribution (that is, scale-free graph). Moreover, since threads in two different blocks cannot communicate\slash synchronize, all auxiliary data structures had to be copied to each thread block. This led to rapid increase in global memory usage if the graph had a large number of nodes. 

The next improvement came from Jia \cite{jia_bc} who improved upon this approach by exploiting neighborhood-level parallelism. Namely, neighbors of a node were explored in parallel. This was done by scheduling threads to each edge instead of to each vertex. This was possible because this algorithm used edge-list data structure as opposed to the adjacency-lists data structure of Sriram's algorithm. However, Jia's algorithm still used one thread block for each source vertex. Although it led to duplication of several large data structures for each thread block, the motivation was that if several thread blocks were to work on the same source vertex, synchronizing and communicating with threads across different blocks would be impossible. Therefore, for both Jia's and Sriram's algorithms, the number of threads that could be deployed for each source vertex was limited due to platform constraints and limitation of global memory. Nevertheless, Jia's algorithm was found to do better than Sriram's algorithm on scale-free networks \cite{jia_bc}.

The BC algorithm by Shi et al. \cite{shi_bc} improved upon Jia's algorithm by making the following observation. Namely, since the data structures must reside in the global memory, there are huge latency involved with each memory reference. They attempted to hide this latency by scheduling a large number of blocks for each source vertex (and thus a fixed number of threads per thread block; this number can be smaller than the maximum permissible threads per block). Whenever a thread block was stalled on memory transaction, it could be replaced by another block. Since different thread blocks were forced to work on the same auxiliary data structures pertaining to a given source vertex, it made sense to have only one copy of these auxiliary data structures. Therefore the memory requirements of this algorithm is smaller  than Jia's algorithm by a factor of $O(n)$, since Jia's algorithm employs $n$ blocks in total. This is a significant achievement since the available global memory on a GPU puts a hard limit on how large a graph can be processed in that device. However, the downside is that in Shi's algorithm different phases in different thread blocks must be synchronized by CPU, the cost which both Jia's and Sriram's algorithm wanted to avoid. However, it was shown that Shi's algorithm achieves considerable speed over Jia's algorithm \cite{shi_bc} for scale-free networks.

In this project, we implemented all three algorithms and observed their performance on various synthetic scale free networks. From these comparisons we drew conclusion on the strength\slash weakness of specific design choices of these algorithms.

%
%\section{Specific Aims}
%In this project, we want to accomplish the following: \begin{enumerate}
%	\item Implement several existing BC algorithms. The most straightforward one is to run a modified APSP algorithm, which in turn runs a modified SSSP from each vertex. The other algorithms are Sriram's algorithm \cite{sriram_bc}, Jia's algorithm \cite{jia_bc}, Shi's algorithm \cite{shi_bc}, and Bader's algorithm \cite{pande_bc_gpu}.
%	\item Compare these algorithms in terms of speed up, memory usage, and resource utilization.
%	\item Identify strengths\slash weaknesses of each algorithm in terms of data structures, thread scheduling, critical region, and network type.
%	\item If possible, improve upon these algorithms by combining strong aspects of different algorithms. For example, it may be possible to reduce the use of locks in Shi's algorithm \cite{shi_bc} by using the idea of lock-free dependency-accumulation of Bader's algorithm \cite{bader_lock_free}.
%	\item Observe the effect of sparse-matrix representation \cite{cusp} for intermediate data structures on the runtime of algorithms.
%\end{enumerate}
%
%

\section{Parallel Programming and Methods}\label{sec:prog}

\subsection{Algorithms}
We implemented three BC algorithms: Shi's algorithm, Jia's algorithm, and lastly Sriram's algorithm. Summary of these algorithms is already discussed in Section~\ref{sec:intro}. Detailed analysis of the data structures used by these algorithms is presented in Section~\ref{sec:mem_req}. In Sriram's algorithm we used shared variables in the BFS kernel to synchronize between different levels. In Shi's algorithm, we used shared memory for a boolean flag variable inside the BFS kernel. Additionally, for Shi's algorithm, we had an additional experimental variable: the number of threads to be scheduled at each thread block in BFS kernel. (For Sriram's and Jia's algorithm, a thread block always used the maximum number of threads allowed.) We verified the correctness of our implementations by comparing the computed BC with the result generated by \texttt{NetworkX} \cite{networkx} -- a graph analysis tool written in Python --  on the same input.

\subsection{Hardware and Software}
Our implementations were written in NVIDIA CUDA C extension. The programs were tested on a high-performance computing cluster, the Hornet cluster at the School of Engineering of the University of Connecticut. The Hornet cluster has 64 nodes, each node with 12 Intel Xeon X5650 Westmere cores and 48 GB of RAM. For of these 64 nodes contain GPUs, each node with eight NVIDIA Tesla M2050 GPUs. Each of these GPUs has 3GB global memory, compute capability 2.0 (which includes floating point atomic operations), 448 thread processors, and maximum 1,024 threads per block (in x-dimension). Our program used only one GPU since the graph was stored entirely in the global memory. Effectively, at most 2,688 MB global memory (out of 3 GB) was available for the graph and all auxiliary data structures.

\subsection{Input Graphs}
We used \texttt{NetworkX} \cite{networkx} to generate synthetic scale-free \textit{directed} graphs using Barabasi-Albert preferential-attachment model. We generated graphs ranging from 500 nodes to 30,000 nodes and textit{Edge-to-Node Ratio}, or \textit{Edge Density} $\displaystyle d=\frac{m}{n}$, from 2 to 60. Here, $n$ and $m$ are the number of vertices and edges, respectively. Note that if the preferential attachment parameter is $\beta$ (that is, each newly added node has $\beta$ \textit{undirected} edges to $\beta$ existing nodes), it follows that $m \approx 2 \beta n$. (The $2$ comes because each undirected edge is replaced with 2 directed edges.)
 
%Moreover, we tried to reduce the memory used in the auxiliary data structures as follows. All these algorithms use a predecessor matrix. In the original description, this matrix was implemented as a boolean array. However, a boolean entry takes 8 bits in the device memory. Therefore we implemented a variant of this data structure using bits. Although this approach reduced the size of the predecessor matrix by a factor of 64, it was necessary that locations in this bit-array should be accessed\slash modified through atomic operations. Thus there was a trade-off between being able to process larger graphs at the cost of additional atomic operations. In our experiments we analyzed whether is trade-off was justified.

\subsection{Memory Usage}\label{sec:mem_req}
The size of the largest graph that could be processed by any BC algorithm is restricted by the amount of available global memory and the size of auxiliary data structures needed by that algorithm. Therefore it is instructive to see how much memory is needed to represent graphs of different size and density by different algorithms. Every algorithm needs the following data structures: \begin{enumerate}
	\item (Input) The graph itself, either in edge list format or in adjacency lists format. In edge list format (adopted by Shi's and Jia's algorithm), there are two arrays $A$ and $B$. For the $i^\mathrm{th}$ directed edge $(u,v)$, we have $A[i]=u$ and $B[i]=v$. Therefore, the size of the graph in edge list format is $2m$, where $m$ is the number of edges in the directed graph. Note that for undirected graphs, each edge can be represented with two directed edges. On the other hand, in adjacency lists format (adopted by Sriram's algorithm), there are two arrays $P$ and $Q$ where $P$ contains one entry for each vertex and $Q$ contains one entry for each edge. Therefore, the size of the graph in adjacency lists format takes $m+n$. Data type for these data structures is integer.
	\item (Output) An array of floating point numbers, one entry per vertex, in total $n$ entries.
	\item (Auxiliary) \begin{enumerate}
		\item (Distance Array) An integer array of distances, one entry per vertex, in total $n$ entries.
		\item (Shortest Path Array) An integer array for keeping track of number of shortest paths through each vertex. In total, it has $n$ entires.
		\item (Dependency Array) An array for accumulating shortest path dependencies for each vertex. In total, it has $n$ floating-point entires.
		\item (Predecessor Array) An array for keeping track of which edge is used to reach each vertex in a shortest path. This array contains $n^2$ entries. The data type, and hence the size, of this array depends on implementation: in some cases it is integer and in other cases it is boolean.
	\end{enumerate}
\end{enumerate}

Below, we examine how much memory each algorithm will need. Note that the size of an \texttt{int} is 4 bytes, the size of a \texttt{float} is 4 bytes, and the size of a \texttt{bool} is 1 byte.

\textbf{Jia's algorithm: }
In Jia's algorithm, one thread block is used for doing BFS from each vertex, and hence the auxiliary data structures are replicated for each thread block. In total there are $n$ concurrent thread blocks launched by the kernel, requiring $n$ replications of the auxiliary data structures. Evidently this approach requires far more global memory -- by several orders of magnitude -- than required by Shi's algorithm. However, Shi's algorithm needs synchronization between successive kernel launches which is avoided by Jia's algorithm by allocating per-block auxiliary data structures. The predecessor array is implemented as $n$ separate arrays, one associated with each source vertex of BFS, each array containing $n$ integers (thus in total $n^2$ boolean entries).

\begin{enumerate}
	\item Input: Edge list, $2m\times 4=8m$ bytes.
	\item Output: $n\times 4=4n$ bytes.
	\item Auxiliary: ($n$ thread blocks combined) \begin{enumerate}
		\item Distance Array: $n\times n\times 4=4n^2$ bytes.
		\item Shortest Path Array: $n\times n\times 4=4n^2$ bytes.
		\item Dependency Array: $n\times n\times 4=4n^2$ bytes.
		\item Predecessor Array:  $n\times n\times 4=4n^2$ bytes.
	\end{enumerate}
\end{enumerate}
Total: $16n^2+8m+4n$ bytes. 

\textbf{Sriram's algorithm: }
The memory footprint of Sriram's algorithm is of the same order as that of Jia's algorithm (data not shown). The difference between these two algorithms (in terms of memory) is that Sriram's algorithm implements graph data structure as adjacency lists, whereas Jia's algorithm uses edge list. The auxiliary data structures used by Sriram's algorithm are the same as those of Jia's algorithm, and like Jia's algorithm, it uses separate thread blocks for doing BFS from each vertex.

\begin{enumerate}
	\item Input: Adjacency lists, $(m+n)\times 4=4m+4n$ bytes.
	\item Output: $n\times 4=4n$ bytes.
	\item Auxiliary: ($n$ thread blocks combined) \begin{enumerate}
		\item Distance Array: $n\times n\times 4=4n^2$ bytes.
		\item Shortest Path Array: $n\times n\times 4=4n^2$ bytes.
		\item Dependency Array: $n\times n\times 4=4n^2$ bytes.
		\item Predecessor Array:  $n\times n\times 4=4n^2$ bytes.
	\end{enumerate}
\end{enumerate}
Total: $16n^2+8n+4m$ bytes.

\textbf{Shi's algorithm: }
In Shi's algorithm,  all data structures are uniformly accessed by all alive thread blocks, and hence the required global memory does not depend on the number of thread blocks used in the algorithm. The predecessor array is implemented as an $n\times n$ boolean array containing $n^2$ entries.

\begin{enumerate}
	\item Input: Edge list, $2m\times 4=8m$ bytes.
	\item Output: $n\times 4=4n$ bytes.
	\item Auxiliary: \begin{enumerate}
		\item Distance Array: $n\times 4=4n$ bytes.
		\item Shortest Path Array: $n\times 4=4n$ bytes.
		\item Dependency Array: $n\times 4=4n$ bytes.
		\item Predecessor Array:  $n^2\times 1=n^2$ bytes.
	\end{enumerate}
\end{enumerate}
Total: $n^2+8m+16n$ bytes.

\subsection{Timing}
We ignored all CPU-level code (e.g., initialization, synchronization, output generation, etc.) of the BC algorithms and timed only the kernel execution using CUDA events.

\section{Results and Analysis}

In this section we examine the memory usage, data structures, thread allocation, speedups, network size and density, and atomic operations. Note that the problem of computing BC, like many other graph-traversing problems, contains less arithmetic computations and more irregular memory accesses. This is why it is hard to optimize the memory reference patterns by writing cache-exploiting code. Also note that all timing data below are presented after averaging over 5 independent runs.

\subsection{Jia's and Sriram's algorithms need high global memory}\label{sec:mem_req_analysis}

\begin{figure}[!h]%
\includegraphics[width=\columnwidth]{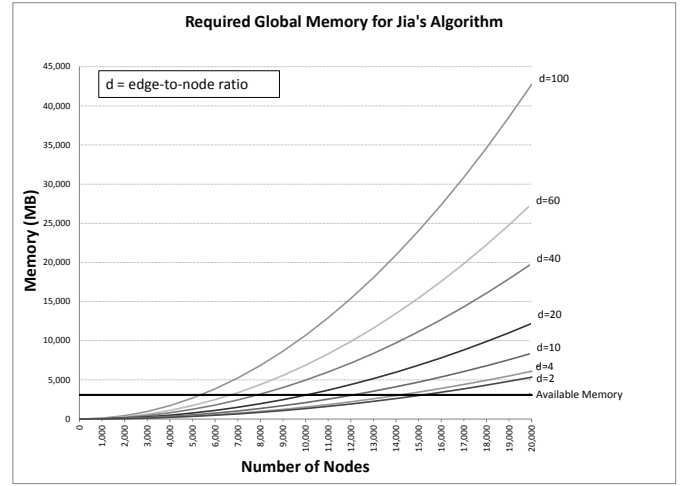}%
\caption{The required global memory for Jia's algorithm as a function of the number of nodes in the input graph. There are 7 plots associated with different $d$ values, where $\displaystyle d=\frac{m}{n}$ is the edge-to-node ratio of the input graph. The thick black horizontal line denotes the maximum available global memory for Tesla M2050 GPU, which is 3 GB.}%
\label{fig:mem_jia}%
\end{figure}

Note that the maximum global memory for a Tesla M2050 GPU is 3 GB. (In practice, we found 2.625 GB available.) Let $\displaystyle d=\frac{m}{n}$ be the edge-to-node ratio of a graph. Figure~\ref{fig:mem_jia} shows how fast the memory requirement of Jia's algorithm grows with the number of vertices in graphs with different densities $d$. This growth is faster for dense graphs. For sparse graphs with $d=2$, the largest graph (in terms of number of nodes) that can be processed by Jia's algorithm given maximum global memory of 3 GB is less than $15,000$ nodes. For denser graphs, the feasible input size is significantly smaller. Memory footprint of Sriram's algorithm is of the same order as that of Jia's algorithm. This huge memory footprint is a major drawback for both Jia's and Sriram's algorithms.

\begin{figure}[!h]%
\includegraphics[width=\columnwidth]{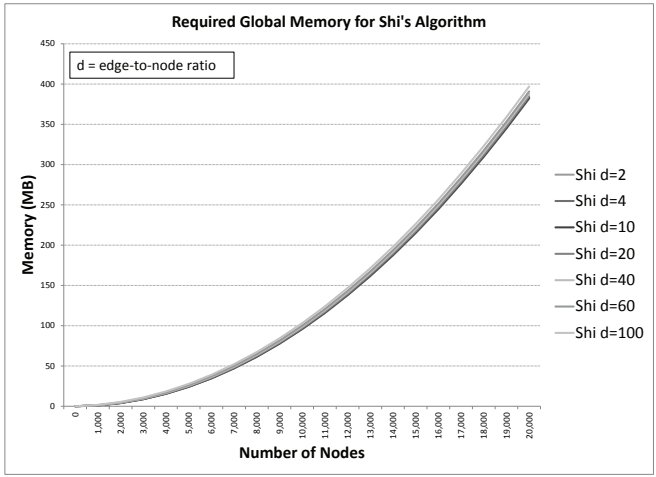}%
\caption{The required global memory for Shi's algorithm as a function of the number of nodes in the input graph. There are 7 plots associated with different $d$ values, where $\displaystyle d=\frac{m}{n}$ is the edge-to-node ratio of the input graph.}%
\label{fig:mem_shi}%
\end{figure}

Figure~\ref{fig:mem_shi} shows the global memory usage of Shi's algorithm. By comparing Figure~\ref{fig:mem_shi} to Figure~\ref{fig:mem_jia}, it can be observed that Shi's algorithm uses much smaller memory than Jia's and Sriram's algorithms. For example, on an input graph with 20,000 vertices and 2,000,000 edges, Shi's algorithm needs 397 MB memory whereas Jia's algorithm needs more than 42 GB memory. Figure~\ref{fig:mem_shi} also shows that in Shi's algorithm, for input graphs with a fixed number of vertices, there is only small variation in memory requirement with respect to edge density. In contrast, the same variation for Jia's algorithm is high (see Figure~\ref{fig:mem_jia}).

\subsection{Shi's algorithm improved as number of nodes increased}

\begin{figure}%
\includegraphics[width=\columnwidth]{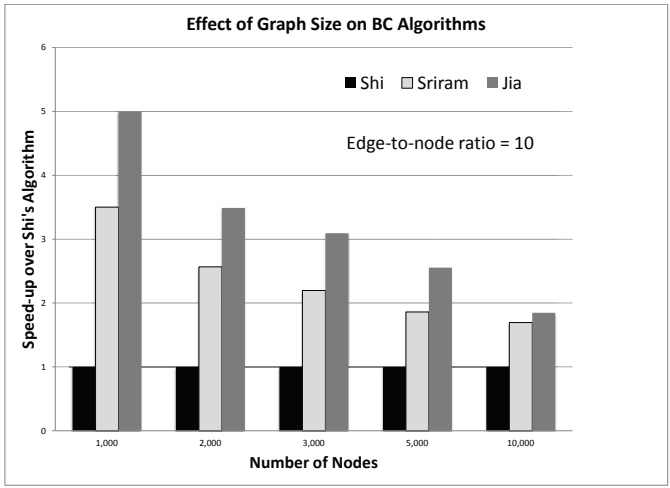}%
\caption{Effect of graph size (with fixed edge density $\displaystyle d=10$) on the BC algorithms. Number of nodes is plotted on the x-axis, and the speed-up of algorithms over Shi's algorithm is plotted on the y-axis.}%
\label{fig:graph_size}%
\end{figure}

It was found that Jia's algorithm was the fastest and Shi's algorithm was the slowest of the three BC algorithms. This fact does not agree with the claim in \cite{shi_bc} which showed that Shi's algorithm was 11\%-19\% faster than Jia's algorithm for input graphs with 10,000-50,000 nodes and different edge densities. As discussed in Section~\ref{sec:mem_req}, we could not test Jia's algorithm on large input graphs due to global memory limitations, and hence could not verify the claim in \cite{shi_bc}. However, Figure~\ref{fig:graph_size} shows that for a fixed edge density, as the number of nodes increased, the speed-up of the other two algorithms over Shi's algorithm decreased. A possible reason for Shi's algorithm not outperforming Jia's algorithm can be the following: the memory latency was not large enough so that scheduling large number of blocks did not help Shi's algorithm. Moreover, the reason Shi's algorithm improves with the increase in the number of nodes is that both Sriram's and Jia's algorithm use separate thread-blocks for each source node in BFS, whereas the number of thread-blocks used by Shi's algoirthm in BFS does not increase so fast with the number of nodes in the input graph (with fixed edge density).

\subsection{In sparse graphs, Jia's algorithm worked better than Sriram's algorithm}

\begin{figure}[!h]%
\includegraphics[width=\columnwidth]{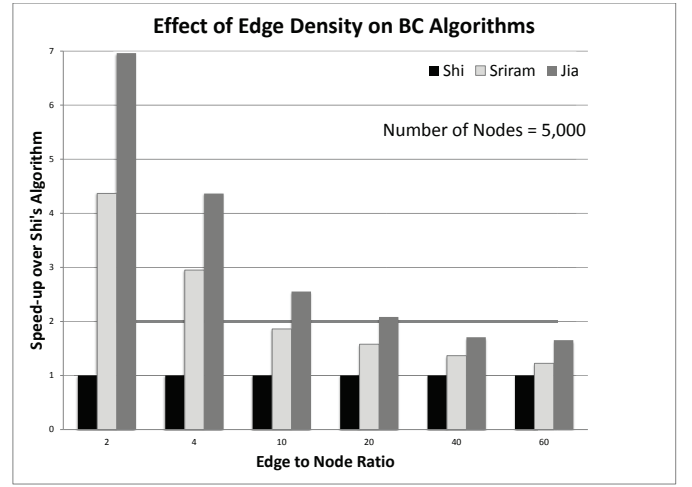}%
\caption{Effect of edge density $\displaystyle d=\frac{\#edges}{\#nodes}$ on the BC algorithms on a graph with 5,000 nodes. Edge density is plotted on the x-axis, and the speed-up of algorithms over Shi's algorithm is plotted on the y-axis.}%
\label{fig:edge_density}%
\end{figure}

Figure~\ref{fig:edge_density} shows that Sriram's algorithm did not do well in sparse graphs (compared to Jia's algorithm). The reason is that it is more likely for Sriram's BFS to encounter a load-imbalanced BFS level (that is, a level containing two vertices one of which has much larger degree than the other) in a sparse network than in a dense network. However, since Jia's algorithm explores edges in BFS instead of vertices, there is no such load-imbalance. As the input graph became more dense, BFS levels with such imbalanced vertices also decreased, and hence the performance disparity between Sriram's and Jia's algorithm also decreased.

\subsection{Shi's algorithm improved for dense graphs}

It can also be observed in Figure~\ref{fig:edge_density} that both Jia's algorithm and Sriram's algorithm were faster than Shi's algorithm. This finding is on the contrary of what claimed in \cite{shi_bc}. Again, one possible reason is that the memory latency was not large enough so that scheduling large number of blocks did not help Shi's algorithm. Additionally, it can also be observed in Figure~\ref{fig:edge_density} that the speed-up of Jia's algorithm over Shi's algorithm decreased as graphs became denser. The reason is,

%\begin{figure}%
%\includegraphics[width=\columnwidth]{speed-up}%
%\caption{Time taken by the kernels of each algorithm (in millisecond) on six networks. The x-axis shows the number of vertices and edges of each network.}%
%\label{fig:speed-up}%
%\end{figure}

\subsection{For Shi's algorithm, scheduling maximum possible threads per block did not yield the best performance}

\begin{figure}%
\includegraphics[width=\columnwidth]{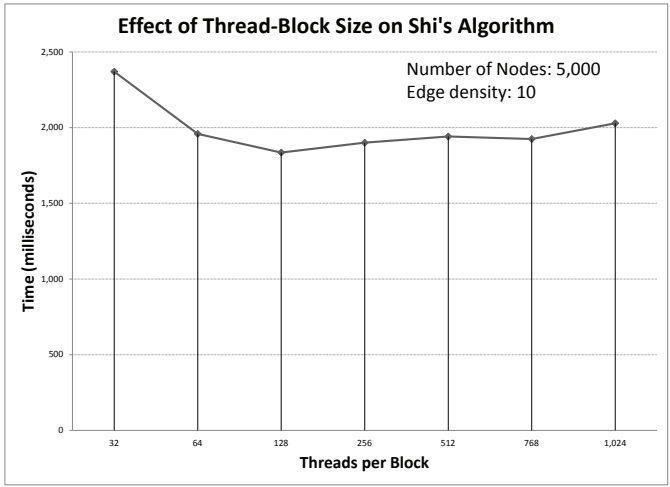}%
\caption{Effect of thread-block size on Shi's algorithm. Number of threads per block is plotted on the x-axis. The execution time (in milliseconds) is plotted on the y-axis. The input graph has 5,000 nodes and 49,950 edges.}%
\label{fig:shi_block}%
\end{figure}

In case of Shi's algorithm, it was observed that scheduling neither maximum (1024) nor minimum (32) number of threads  per block yielded the best performance. The best performance of the algorithm usually came when a smaller number of threads were scheduled per block. Figure~\ref{fig:shi_block} shows that for an input graph with 5,000 vertices and 49,950 edges, the best performance was achieved when 128 threads were scheduled per block. This happened because scheduling too few threads per block is a waste of compute capability, since GPU multiprocessors can have only a fixed number of active blocks at any time. On the other had, whenever a thread stalls for global memory reference, all other threads in that block have to wait. This suggests that if there is possibility of many random, non-cache memory references (as in the case of computing BC), scheduling the maximum number of threads per block may lead to under-utilization of resources.

\subsection{Our implementation of Shi's algorithm performed better than the original implementation}\label{sec:shi-orig}
As claimed in \cite{shi_bc}, Shi's algorithm was at least 10\% faster than Jia's algorithm on graphs with 10,000 -- 50,000 vertices and having edge density $10\leq d \leq 50$. However, we could not reproduce this claim in our experiments; in our experiments, Jia's algorithm (and also Sriram's algorithm) always performed better than Shi's algorithm. It should be noted that due to global memory limitations, we could test Jia's algorithm on graphs with much fewer size and edge densities (see Figure~\ref{fig:mem_jia}). However, a pattern can be observed in Figure~\ref{fig:graph_size} and Figure~\ref{fig:edge_density} that as input graphs became larger\slash more dense, the speed-up of Jia's algorithm over our implementation of Shi's algorithm became smaller. If this trend continues for larger\slash denser graphs, at some point Shi's algorithm will perform better than Jia's algorithm.

\begin{figure}%
\includegraphics[width=\columnwidth]{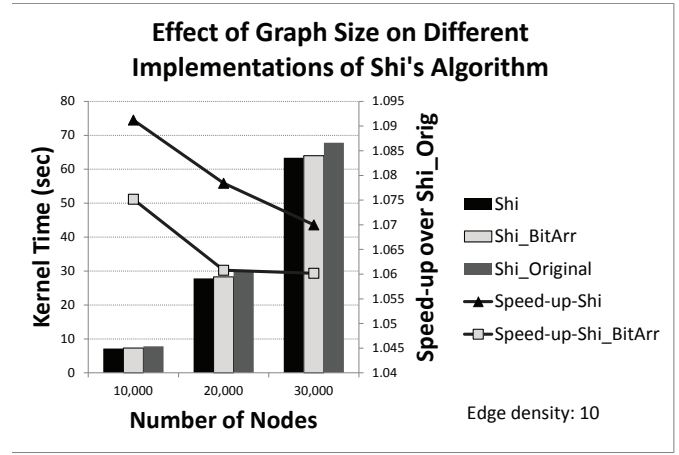}%
\caption{Comparison between our implementation of Shi's algorithm (both with\slash without the bit-array predecessor matrix) against the original implementation on input graphs with fixed edge density $d=10$. The x-axis shows number of nodes. The primary y-axis (left) shows the execution time, and the secondary y-axis (right) shows the speed-up of our implementation (both with\slash without the bit-array predecessor matrix) over the original implementation. Number of threads scheduled per block for BFS was 1,024 (maximum).}%
\label{fig:shi-orig-graph-size}%
\end{figure}

\begin{figure}%
\includegraphics[width=\columnwidth]{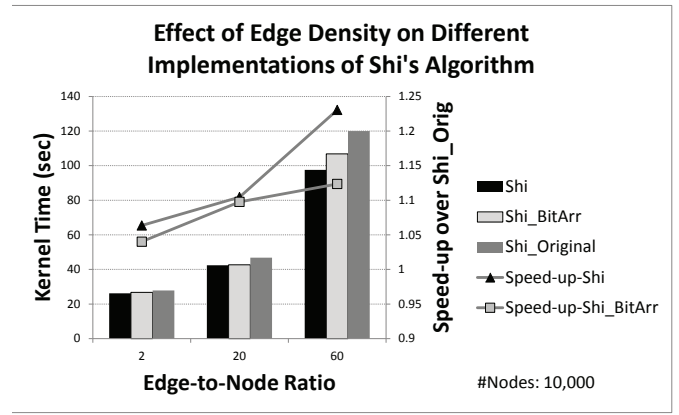}%
\caption{Comparison between our implementation of Shi's algorithm (both with\slash without the bit-array predecessor matrix) against the original implementation on input graphs with 10,000 vertices. The x-axis shows number of nodes. The primary y-axis (left) shows the execution time, and the secondary y-axis (right) shows the speed-up of our implementation (both with\slash without the bit-array predecessor matrix) over the original implementation. Number of threads scheduled per block for BFS was 1,024 (maximum).}%
\label{fig:shi-orig-edge-density}%
\end{figure}

Due to this discrepancy in the performance of Shi's algorithm, we compared the original implementation of Shi's algorithm against our implementation. We downloaded the source code mentioned in \cite{shi_bc} and used the BC procedure from our code. We call this \textit{the original Shi algorithm}. Figure~\ref{fig:shi-orig-graph-size} and Figure~\ref{fig:shi-orig-edge-density} show that our implementation always performed better than the original implementation, with speed-up ranging from 6\%-22\%. Figure~\ref{fig:shi-orig-graph-size} shows that our implementation yielded higher speed-up when the input graph became smaller. Likewise, Figure~\ref{fig:shi-orig-edge-density} shows that our implementation yielded higher speed-up when the input graph became more dense.

Since the original implementation of Shi's algorithm never performed better than our own implementation, it follows that the original Shi's algorithm would not perform better than Jia's algorithm because Jia's algorithm was always faster than our Shi's algorithm on the feasible input graphs within the memory limitation (maximum 3 GB global memory, maximum 10,000 vertices, maximum edge density 20). Thus we could not validate the claim of \cite{shi_bc} (that Shi's algorithm is superior to Jia's algorithm) within our memory limitations.

\subsection{Code optimization yielded improvement over original implementation of Shi's algorithm}\label{sec:shi-improvement}

Two reasons why our implementation of Shi's algorithm performed better than the original implementation (see Section~\ref{sec:shi-orig}) are the following.
\begin{enumerate}
	\item \textit{Code optimization} Inside the BFS thread-block, there is a boolean variable which marks whether there is another level to explore. This memory location is referenced from different threads from different blocks in parallel.In the original implementation this variable was created in the global memory whereas in our implementation it was created in shared memory. This optimization reduced many unnecessary global memory references. Figure~\ref{fig:shi-orig-graph-size} and Figure~\ref{fig:shi-orig-edge-density} show that our implementation with a bit-array predecessor matrix was faster than the original implementation. This proves that the optimization (using shared memory flag variable as described above) indeed improved the performance of Shi's algorithm.
	
	Moreover, since the fraction of BFS levels (with respect to the total number of nodes) becomes higher as the graph becomes more dense (assuming fixed number of nodes), the performance gain from this optimization in our implementation of Shi's algorithm over the original implementation will be more evident in graphs with higher edge density. This phenomenon is evident in Figure~\ref{fig:shi-orig-edge-density}. 
	
		When the size of the input graph grows without changing the edge density, individual neighborhoods (that is, subgraphs) still \textit{looks} the same (in terms of degree distribution). Thus with increase in size (but fixed edge density), the fraction of BFS levels (with respect to the number of nodes in the graph) tends to decrease. Hence the fraction of lookups\slash updates to the abovementioned flag variable also decreases. This explains why in Figure~\ref{fig:shi-orig-graph-size} the speed-up due to this code optimization decreased as the size of the graph increased (from left to right).
	
	\item \textit{Avoiding atomic operations} Note that the predecessor matrix of Shi's algorithm (see Section~\ref{sec:mem_req}) contains $n^2$ boolean entries. In the original implementation, it was implemented as a bit-array so that each entry was represented by a single bit whereas in our implementation, it was implemented as a regular $n\times n$ boolean array. The reason of using a bit-array in the original implementation was to save memory: it reduced the size of the predecessor matrix by a factor of 64. However, the price to pay was that now every update to this bit-array had to be atomic. Since atomic operations to the global memory is costly, it led to performance degradation of the original implementation compared to our implementation which uses slightly more memory but did not need atomic operations to modify the predecessor matrix. This is why in both Figure~\ref{fig:shi-orig-graph-size} and Figure~\ref{fig:shi-orig-edge-density} our implementation without the bit-array predecessor matrix was always faster than those using bit-array predecessor matrix. 
	
	Additionally, as the input graph became more dense (see Figure~\ref{fig:shi-orig-edge-density}), there were more BFS levels and more udpates to the predecessor matrix; this is why in Figure~\ref{fig:shi-orig-edge-density} the speed-up of our implementation without bit-array predecessor matrix became higher than the speed-up of our implementation with bit-array predecessor matrix as the graph became more dense (from left to right).
	\end{enumerate}
	
\subsection{Shi's algorithm had large data transfer time} Although Jia's and Sriram's algorithm uses larger memory than Shi's algorithm, the latter allocates memory for each BFS level whereas the former two (Jia's and Sriram's) algorithms allocate data only once. This aspect of Shi's algorithm adds a significant amount of time for data transfer (data not shown) which we did not report in timing\slash speed-up comparisons. The reason was, time for memory allocation depends on many things (e.g., PCI bus, sequence of requests, etc.) and cannot be faithfully reproduced.

\section{Critiques}

In this work we tried to effectively compare various aspects of different BC algorithms on GPU. However, our approach can be improved by addressing the following issues. 

\textit{Timing} we could not accurately measure the actual time necessary for memory allocation and copying since these calls involve communication over PCI bus as well as invocation of various device level procedures (for example, CUDA context for the first CUDA call) which is hard to reproduce consistently. Therefore, the time used in our comparisons is only the kernel execution time and not the time actually experienced by the user (i.e., the wall clock time).

\textit{Data types} The default size of \texttt{int} is 32 bits. However, the graphs used in this study had at most 30,000 nodes, and therefore the node numbers could be represented using \texttt{int16}. This would reduce the size of auxiliary data structures by a factor close to 2.

\textit{Discrepancy in Shi's algorithm} The performance of Shi's algorithm reported in \cite{shi_bc} could not be verified because the auxiliary data structures for Jia's algorithm would be too large (according to the memory requirement discussed in Section~\ref{sec:mem_req}), which undermines the study.

\textit{Device utilization} We did not present an estimate of the utilization of the device as is found on the literature: usual FLOPS or edge traversed per second \cite{bader_lock_free}. Thus it remains unclear how much each design choice affected the utilization.

\textit{Real-world input} No real world networks was used in this study. Moreover, only scale-free networks have been used.

\textit{Thread-block size for Shi's algorithm} It would be better if some suggestion could be made about the optimal number of threads per block for the BFS kernel of Shi's algorithm. Currently, only one case (a graph with 5,000 vertices and edge density 10) is presented. More cases with variation in the graph size and edge density might reveal such a pattern.

\textit{Some unanswered questions} There remains some unanswered questions. For example, the explanations for the variation in speed-ups (in Figure~\ref{fig:shi-orig-graph-size} and Figure~\ref{fig:shi-orig-edge-density}) due to the code optimization (see Section~\ref{sec:shi-improvement}) were not substantiated with experimental evidence (e.g., counting the number of BFS levels at each input graph). Also, in those two figures, there were only three data points; it would be better if thre were more data points.

\section{Future Works}
We can suggest the following future works from our study: \begin{enumerate}
	\item Use shared variables in CUDA kernel functions of all BC algorithms to improve performance. This might make the code more complicated, but this aspect is worth investigating.
	\item Use different representation methods (e.g., sparse matrix methods) for auxiliary data structures. However, these methods must not introduce additional atomic operations.
	\item Adaptively select integer data types (e.g., \texttt{int32} or \texttt{int16}) for auxiliary data structures by looking into the meta-information (e.g., number of nodes) about the network. That way the global memory requirement would be reduced and larger input graphs would be feasible for analysis.
	\item Investigate ways to analyse large graphs that do not fit into the global memory of a single GPU.
\end{enumerate}

\section{Conclusions}

In this project we implemented three algorithms for computing betweenness centrality on the GPU, namely Sriram's algorithm \cite{sriram_bc}, Jia's algorithm \cite{jia_bc}, and Shi's algorithm \cite{shi_bc}. We showed how Jia's algorithm outperformed Sriram's algorithm by utilizing edge-level parallelism for exploring neighborhoods. However, our results about Shi's algorithm did not conform to the performance reported by the authors in \cite{shi_bc}. Moreover, we showed that making a simple code optimization inside the BFS kernel of Shi's algorithm yields better performance than the original implementation.

\bibliographystyle{IEEEtran}
% argument is your BibTeX string definitions and bibliography database(s)
%\bibliography{../../../../Citations/Mendeley/bc}
\bibliography{bc}
%
% <OR> manually copy in the resultant .bbl file
% set second argument of \begin to the number of references
% (used to reserve space for the reference number labels box)
%\begin{thebibliography}{1}

%\end{thebibliography}

% that's all folks
\end{document}